\documentstyle[aps,epsf]{revtex}
\newcommand{\be}{\begin{equation}}
\newcommand{\ee}{\end{equation}}
\newcommand{\bea}{\begin{eqnarray}}
\newcommand{\eea}{\end{eqnarray}}

\newcommand{\eqn}[1]{(\ref{#1})}

\def\sac{\, , \qquad}

\newcommand{\ra}{\rightarrow}
\newcommand{\calf}{\mbox{${\cal F}$}}
\newcommand{\cale}{\mbox{${\cal E}$}}
\newcommand{\calb}{\mbox{${\cal B}$}}

\newcommand{\adss}[2]{$AdS_{#1} \times S^{#2}$}
\newcommand{\w}{\wedge}

\newcommand{\e}{\epsilon}
\newcommand{\fc}{\frac}
\renewcommand{\a}{\alpha}
\renewcommand{\d}{\delta}

\newcommand{\ik}{{\it k}}

\newcommand{\itwo}{{\it 2}}
\newcommand{\ithree}{{\it 3}}

\newcommand{\jhep}[3]{{J. High Energy Phys.} {\bf #1} {(#2)} #3}

\newcommand{\npb}[3]{{Nucl. Phys.} {\bf B #1} {(#2)} #3}
\newcommand{\plb}[3]{{Phys. Lett.} {\bf B #1} {(#2)} #3}

\renewcommand{\prl}[3]{{Phys. Rev. Lett.} {\bf #1} {(#2)} #3}

\newcommand{\cqg}[3]{{Class. and Quant. Grav.} {\bf #1} {(#2)} #3}

\newcommand{\hepth}[1]{{\tt hep-th/#1}}

\begin{document}

\twocolumn[\hsize\textwidth\columnwidth\hsize\csname
@twocolumnfalse\endcsname

\rightline{DAMTP-2002-78}
\rightline{hep-th/0206194}

\title{Penrose Limits, Worldvolume Fluxes and Supersymmetry} 
\author{David Mateos\\}
\address{
DAMTP, Centre for Mathematical Sciences, University of Cambridge \\
Wilberforce Road, Cambridge CB3 0WA,  UK \\
{\small D.Mateos@damtp.cam.ac.uk} \\
}
\maketitle
\begin{abstract}
It is noted that a finite Penrose limit for brane probes with 
non-zero worldvolume fluxes does {\it not} generically exist; 
this is closely related to the observation by Blau et al that 
for a brane probe the Penrose limit is equivalent to an 
infinite-tension limit. It is shown that when the limit exists,
however, the number of supersymmetries preserved {\it by the probe} does 
not decrease.

\end{abstract}
\vskip2pc]

\section{Introduction}

Three crucial features of the Penrose limit \cite{Penrose76}
of a supergravity solution \cite{Guven00} along a null geodesic are 
({\it i}) that a finite limit always exists, ({\it ii}) that 
the resulting configuration is also a solution of the supergravity 
field equations, and ({\it iii}) that the number of supersymmetries 
of the initial solution does not decrease in the limit \cite{BFP02}. 
In fact, the precise rescalings of the supergravity fields involved 
in taking the limit are dictated by demanding that conditions 
({\it i}) and ({\it ii}) be satisfied.

Brane probes in supergravity spacetimes\cite{probe}
are an essential tool for understanding both spacetime geometry and 
gauge theory physics \cite{Johnson00}.  
One purpose of this letter is to observe that property ({\it i}) 
does {\it not} extend to solutions of the equations of motion
of a brane probe with {\it generic} non-zero worldvolume gauge fields; 
this will be illustrated with a simple example. This result is
closely related to the fact that the Penrose limit for the
supergravity background can be reinterpreted as an infinite-tension 
limit for the probe \cite{BFHP02,BFP02}; we will return to this 
in the Discussion.

In the context of the AdS/CFT correspondence, two examples of 
brane probes in \adss{5}{5} with non-zero worldvolume gauge fields 
are provided by the `defect' D5-brane \cite{KR01} and 
the `baryonic' D5-brane \cite{Witten98}. These are especially relevant 
to the present discussion because, as we will explain, they share a
non-generic feature that allows the general arguments presented here to
be circumvented and thus a finite limit to be defined.

Property ({\it ii}) does extend to brane probe solutions
\cite{BFHP02,BFP02} provided of course a well-defined Penrose 
limit exists. The introduction of the probe generically preserves only
a fraction (possibly none) of the supersymmetries of the
supergravity background. It has been tacitly assumed in the literature 
that the number of supersymmetries preserved {\it by the probe} 
does not decrease in the Penrose limit, that is, that property 
({\it iii}) also extends to brane probes. Since to our 
knowledge no proof of this has been presented, we provide a simple one
here.

\section{Worldvolume Fluxes}
\label{fluxes}

In the neighbourhood of a segment of a null geodesic with no 
conjugate points the metric may be written as 
\cite{Penrose76,Guven00,BFP02}
\be
g = dV \left( dU + \alpha \, dV + \beta_i \, dY^i \right) +
C_{ij} \, dY^i dY^j \,,
\ee
where $\alpha$, $\beta_i$ and $C_{ij}$ are functions of all the coordinates. 
The geodesic lies at $V=Y^i=0$ and is affinely parametrized by
$U$. Let us introduce new rescaled coordinates 
\be
U=u \sac V=\Omega^2 v \sac Y^i = \Omega y^i \,, 
\ee
where $\Omega$ is a positive real constant, and set 
$g_\Omega =\Omega^{-2} g$. The Penrose limit of $g$ is obtained by 
computing the limit 
\be
\bar{g} \equiv \lim_{\Omega \ra 0} g_\Omega
\label{g-limit}
\ee
while keeping $u$, $v$ and $y^i$ fixed. Similarly, if $B_{\ik-1}$ 
is a supergravity gauge potential subject to the gauge transformation
\be
\d B_{\ik-1} = d \Lambda_{\ik -2}
\label{B-trans}
\ee
and $H_{\ik}=d B_{\ik-1}$ is its gauge-invariant $k$-form field
strength\cite{complicated}, then we set 
\be
H_\Omega = \Omega^{-k+1} H
\label{H-resc}
\ee
and define the Penrose limit of $H$ as 
$\bar{H} \equiv \lim_{\Omega \ra 0} H_\Omega$, again keeping $u$, $v$
and $y^i$ fixed. It is easy to see that these limits are 
finite \cite{Penrose76,Guven00,BFP02}. 

In the presence of branes there are in general additional
gauge potentials $A_{\ik-1}$ (for certain values of $k$) that describe 
degrees of freedom localized on the worldvolumes of the branes.
Examples of these include a one-form gauge potential in the case 
of D-branes and a two-form gauge potential in the case of the 
M5-brane. Their gauge-invariant field strengths $\calf_k$ involve the 
supergravity gauge potentials through combinations of the form 
\be
\calf_{\it k} = F_{\it k} + B_{\it k}^\star \,,
\label{combi}
\ee
where $F_{\it k} = d A_{\ik -1}$ and `$\star$' denotes 
the pull-back to the worldvolume. 
The reason is that in the presence of branes the theory is not
invariant under the gauge transformations \eqn{B-trans} alone but in
combination with\cite{gauge}
$\d A_{\it k-2} = - \Lambda_{\it k-2}^\star$.
In the case of D-branes $B_\itwo$ is the Neveu-Schwarz two-form, 
whereas for the M5-brane $B_\ithree$ is the three-form potential 
of eleven-dimensional supergravity. 

Equation \eqn{combi} has crucial implications for the existence of  
finite Penrose limits for brane probes because it means 
that, in order to extend the definition of the limit to the 
worldvolume gauge fields in a {\it gauge-invariant} manner, we must
define $F_\Omega = \Omega^{-k} F$ and  
$\bar{F} \equiv \lim_{\Omega \ra 0} F_\Omega$. It then follows
that $\calf_\Omega = \Omega^{-k} \calf$ and 
\be
\bar{\calf} \equiv \lim_{\Omega \ra 0} \calf_\Omega \,.
\label{calf-limit}
\ee
We thus see that gauge invariance forces the $k$-form 
field strength $\calf_\ik$ to be rescaled with one extra power 
of $\Omega^{-1}$ as compared to a supergravity field strength of 
the same rank. 

To see the consequences of this, consider a null geodesic that
intersects, or is contained within\cite{close}, the worldvolume of the brane. 
For {\it generic} non-zero values of $F_\ik$ and $B_\ik$ the components 
$\calf_{U i_1 \ldots i_{\ik-1}}$ of the worldvolume flux $\calf_\ik$
will be non-zero in a neighbourhood of the geodesic. 
Since these components lead to terms that scale as
$\Omega^{k-1}$, the limit \eqn{calf-limit} diverges. Needless to say,
one may choose to gauge away the analogous components of $B_\ik$ or of
$F_\ik$, but the gauge-invariant field strength $\calf_\ik$ will
remain unchanged. 

Since $\calf_\ik$ has support solely on the worldvolume of the brane,
it is clear that the presence of the probe does not affect the Penrose limit
along a null geodesic that lies entirely ouside the worldvolume.
In fact, since only a region infinitesimally 
close to the geodesic (which is magnified by an infinite amount) 
survives the limit, the brane just disappears from 
the resulting spacetime. 

The obstruction to the existence of a finite Penrose limit is not
visible in an effective description in which the brane is replaced 
by its backreaction on spacetime, that is, by some supergravity
solution, because in this description the worldvolume flux $\calf$ 
is ignored\cite{ignored}.
Moreover, most of these supergravity solutions are typically 
singular at the location of the putative brane, so their Penrose
limits along geodesics that intersect the brane are not usually considered. 

The brane probe action is homogeneous under the rescalings
above \cite{BFHP02,BFP02}; for example, the action for a 
D$p$-brane in a supergravity background
\be
S_p = - \int e^{-\phi} \sqrt{-\det(g+\calf)} + 
\int e^{\calf} \w C 
\label{action}
\ee
is homogeneous\cite{dilaton} of degree $-(p+1)$. 
This property ensures that 
solutions of the brane equations of motion in a supergravity spacetime 
are mapped by the Penrose limit to new solutions in the 
resulting spacetime, provided of course that the limit for the
probe exists. 

Brane probes play an important role in the context of the AdS/CFT
correspondence. Two examples in an \adss{5}{5} background are
the defect D5-brane \cite{KR01} and the baryonic D5-brane 
\cite{Witten98}. These are especially relevant to 
the present discussion because the worldvolume flux $\calf$ is non-zero 
in both cases, yet finite Penrose limits for the defect brane and for
the baryonic brane were found in  
\cite{ST02} and \cite{MN02}, respectively \cite{clarification}.
The non-generic feature common to these two cases that allows the
arguments above to be circumvented is that the overall 
scale of $\calf$ is {\it arbitrary}; more precisely, 
there is an entire family of solutions parametrized by the magnitude 
of $\calf$ \cite{precise}.
This implies that $\calf$ can be rescaled with an arbitrary power of
$\Omega$ that, in particular, can be appropriately chosen 
in order to make the limit \eqn{calf-limit} finite; 
in effect, this means that one does not take the Penrose limit 
of a fixed solution, but instead focuses on an $\Omega$-dependent
member of the family as $\Omega$ is scaled to zero.
It is this additional freedom that allowed a finite limiting result to 
be obtained in \cite{ST02,MN02}.

\section{An Example}
\label{example}

A simple example in which the freedom discussed above 
to rescale the worldvolume flux at will does not occur is provided by the
supersymmetric D3/$\overline{\mbox{D3}}$ system. 
This consists of a stack of $N$ 
infinite flat D3-branes separated by some distance from a
parallel stack of $\bar{N}$ anti-D3-branes. On the worldvolumes of
both groups of branes there are constant (abelian) electric fields $\cale$ 
and $\cale'$ (aligned with each other) and constant magnetic 
fields $\calb$ and $\calb'$ (also aligned with each other but
orthogonal to the electric fields). 
For generic values of these worldvolume fluxes this system is unstable: 
there is a long-distance force between the two groups 
of branes \cite{force}, and for sufficiently small separation an open-string 
tachyonic mode develops. However, if $\cale=\cale'=\pm 1$ and $\calb$ 
and $\calb'$ are non-zero \cite{subcritical}
and have opposite signs, then the whole system preserves 1/4 
supersymmetry (hence the force vanishes \cite{EMT01,BL01} and no 
tachyonic instability appears \cite{MNT01,BO01}). 
This follows from the fact that this configuration is T-dual to 
the supersymmetric D2/$\overline{\mbox{D2}}$ system \cite{BK01,MNT01}, 
whose supersymmetry can in turn be understood from its origin 
as a particular limit of the D2-brane supertube \cite{MT01}.

If $N=1$ and $g_s \bar{N} \gg 1$, where $g_s$ is the string coupling
constant, then the appropriate description of this
system is in terms of a D3-brane probe in the supergravity background
created by the $\bar{N}$ $\overline{\mbox{D3}}$-branes. 
In this case the condition 
$|\cale|=1$ arises from the requirement that the equations of motion 
derived from the action \eqn{action} with $p=3$ are solved by a static 
D3-brane probe with worldvolume electric and magnetic fields 
\cite{EMT01}. 
(Note that although this condition happens to imply preservation 
of 1/4 supersymmetry, this need not be imposed a priori.) 
Since the value of the worldvolume electric field on the probe
is now {\it fixed}, it follows from the arguments in the 
previous section that the Penrose limit along {\it generic} geodesics 
that intersect (or are contained within) the D3-brane leads
to a divergent result for the worldvolume field strength.

\section{Supersymmetry}

The Penrose limit of a supergravity solution possesses at least as
many supersymmetries as the original solution\cite{BFP02}. 
The essence of the argument is as follows. For each Killing spinor 
$\e$ of the initial solution there exists a real constant $\beta$ 
such that the limit 
$\bar{\e} \equiv \lim_{\Omega \ra 0} \Omega^{\beta} \e$
is finite. The linearity of the Killing spinor equations then 
implies that $\bar{\e}$ is a Killing spinor of the resulting
supergravity solution. The argument is completed by showing that it is
possible to choose the initial basis of Killing spinors in such a way
that the limiting ones are linearly independent.

The supersymmetries of a supergravity solution that are left unbroken
by the introduction of a brane probe are \cite{BKOP97}
those generated by spacetime Killing spinors $\e$ that satisfy 
\be
\Xi \, \e = \sqrt{-\det(g+\calf)} \, \e \,,
\label{kappa}
\ee
where $\Xi$ is the matrix appearing in the kappa-symmetry
transformations of the brane worldvolume fermions; by
construction it satisfies $\Xi^2 = -\det(g+\calf)$. 
Equation \eqn{kappa} may be regarded as the worldvolume analogue of the
background Killing spinor equations. The relevant observation for our
purpose is that \eqn{kappa} is {\it linear} and that both sides 
are homogeneous of the {\it same} degree under the rescalings 
involved in the Penrose limit; we show this explicitly below 
for D-branes. It follows that if a
subset of the background Killing spinors $\{\e_i\}$ verify 
\eqn{kappa} then so do their Penrose limits $\{\bar{\e_i}\}$, 
and therefore the resulting probe will preserve at least as many 
supersymmetries as the original one. Of course, just like in certain
cases \cite{enhancement} the number of supersymmetries of the
background may actually increase in the limit, so may do the number of
supersymmetries preseved by the brane \cite{ST02,MN02}.

One important feature of equation
\eqn{kappa} is \cite{BT98} that it is well-defined if 
$\det g=0$ and/or $\det (g+\calf)=0$ (in the latter case $\Xi$ is 
nilpotent), as may be the case for the resulting brane 
in the Penrose limit even if it was not the case for the original
brane \cite{MN02}.  

The kappa-symmetry matrix for D$p$-branes can be found in
\cite{BT98}; since the formulas differ slightly in the type IIA and
type IIB cases, we focus here on type IIB D$p$-branes for concreteness. 
In this case $\Xi$ is defined by the equation 
\be
\Xi \, d\sigma^0 \w \ldots \w d\sigma^p = 
\sum_n \Gamma_{(2n)}^\star \, K^n I \w e^{\calf} \,,
\label{xi}
\ee
where $\{\sigma^0, \ldots ,\sigma^p\}$ are worldvolume coordinates
in which the determinant on the right hand side of \eqn{kappa} is
calculated. $\Gamma_{(n)}^\star$ is the pull-back to the worldvolume 
of the spacetime matrix-valued $n$-form
\be
\Gamma_{(n)} = \fc{1}{n!} \Gamma_{a_1 \ldots a_n} 
e^{a_1} \w \ldots \w e^{a_n} \,,
\ee
where $\{\Gamma_a\}$ are ten tangent-space constant Dirac matrices
and $\{e^a\}$ is a basis of orthonormal one-forms for the
spacetime metric, that is, $g= \eta_{ab} \, e^a e^b$. 
$K$ and $I$ are linear operators that act on 16-component 
complex spinors of type IIB supergravity as $K \psi = \psi^*$ and
$I\psi =-i \psi$. Finally, it must be understood that only the form 
of degree $p+1$ is selected on the right hand side of \eqn{xi}. 

The conformal rescaling of the background metric involved in the
Penrose limit implies that the orthonormal one-forms are rescaled as 
$e^a_\Omega = \Omega^{-1} e^a$. This, together with the
rescaling of $\calf$, implies that $\Xi_\Omega = \Omega^{-p-1} \Xi$.
It then follows that both sides of \eqn{kappa} are homogeneous 
of the same degree $\beta -p -1$.

\section{Discussion}

In the preceding discussion the tension of the brane probe was
set to unity, and therefore implicitly kept {\it fixed} in 
all the rescalings involved in the Penrose limit. 
It was shown in \cite{BFHP02,BFP02},
however, that the Penrose limit of the supergravity background 
can be reinterpreted as an infinite-tension limit for the probe. 
Let us see the connection between this result and the definition of the
Penrose limit that we have adopted. For concreteness, we focus again
on a D$p$-brane. If we reinstate $\a'$ then the gauge-invariant field 
strength \eqn{combi} becomes
\be
\calf = \a' F + B^\star \,.
\label{calf-a'}
\ee
In addition, the action \eqn{action} acquires an overall factor of 
$(\a')^{-\frac{p+1}{2}}$, and therefore it satisfies \cite{BFHP02,BFP02} 
\be
S_p [\Psi_\Omega, F; \a'] = S_p [\Psi, F; \tilde{\a}'] \,,
\label{formula}
\ee
where $\Psi$ collectively denotes all the background fields and 
$\tilde{\a}' = \Omega^2 \a'$. 
We thus see that the Penrose limit $\Omega \ra 0$ 
for the background spacetime
translates into an infinite-tension limit $\tilde{\a}' \ra 0$ 
for the brane in the original background. 
On the other hand, if we now rewrite equation \eqn{formula} as 
\be
S_p [\Psi_\Omega, F; \Omega^{-2} \a'] = S_p [\Psi, F; \a'] 
\ee
then we may reinterpret it as saying that the D$p$-brane action is
homogeneous (of degree zero) if the background fields are rescaled as
above, $F$ is kept fixed and $\a'$ is rescaled as 
$\a'_\Omega = \Omega^{-2} \a'$. Note that this definition of the limit
is consistent with gauge invariance because both terms on the right
hand side of \eqn{calf-a'} are rescaled in the same way. 
In fact, as far as the Penrose limit of {\it classical} 
solutions of the action
\eqn{action} is concerned, this definition and that of section
\ref{fluxes} always yield {\it physically equivalent} results (in
particular, they both fail to exist in the same cases). The reason is
that, except for the overall factor in the action \eqn{action}, 
$\a'$ and $F$ do not occur separately but only in the combination 
$\a' F$, which in both cases is rescaled in the same way. 
The different overall scalings (degrees of homogeneity)
of the action in the two cases could make a difference 
when this is inserted in some path-integral, but they do not make 
a difference at the classical level. 

In this letter we have focused on the obstructions to the existence of
the Penrose limit associated to the presence of worldvolume
fluxes. In some cases, however, the existence of a well-defined limit 
for the brane {\it embedding} itself is not a trivial issue
\cite{MN02}. This is consistent with the fact that the embedding is
specified by worldvolume scalar fields and that these may be dual to
worldvolume gauge fields in certain cases. 
One example of this is provided by the type
IIA D2-brane: its worldvolume one-form potential is equivalent to a
periodically-identified scalar field which, after the reinterpretation
of the D2-brane as an M-theory membrane, specifies the position of the 
latter along the M-theory circle \cite{BT96}. 
It follows that any obstacles to the existence of
the Penrose limit due to the D2-brane gauge field must be reinterpretable 
in terms of the M2-brane embedding.

\medskip
\section*{Acknowledgments}
\noindent
It is a pleasure to thank Jos\'e Figueroa-O'Farrill, Chris Hull
and Selena Ng for helpful discussions and comments on the manuscript. 
This work was supported by a PPARC fellowship.



\begin{thebibliography}{99}
\bibitem{probe}
Brane probes are branes whose backreaction on spacetime can be
neglected.
\bibitem{close}
Or that becomes arbitrarily close to the brane; for a
discussion of this case see \cite{MN02}.
\bibitem{dilaton}
The dilaton is not rescaled in the Penrose limit.
\bibitem{precise}
The precise way in which the brane is embedded depends on 
the magnitude of $\calf$.
\bibitem{force}
This force may be attractive or repulsive depending on the values and
signs of the electromagnetic fields; in the particular case in which
they all vanish the force is attractive.
\end{thebibliography}
\end{document}